# Influence of Nb Alloying on Nb Recrystallization and the Upper Critical Field of Nb$_3$Sn

Nawaraj Paudel[1], Chiara Tarantini[1], Shreyas Balachandran[1,†], William L. Starch[1],
Peter J. Lee[1], David C. Larbalestier[1,*]

[1]*Applied Superconductivity Center, National High Magnetic Laboratory, Florida State University, Tallahassee, FL 32310, USA*
[†]*Now at Jefferson Laboratory, Newport News, VA 23606, USA*

Nb$_3$Sn conductors are important candidates for high-field magnets for particle accelerators, and they continue to be widely used for many laboratory and NMR magnets. However, the critical current density, $J_c$, of present Nb$_3$Sn conductors declines swiftly above 12-15 T. State-of-the-art Ta- and Ti-doped strands exhibit upper critical field, $H_{c2}$, values of ~ 24-26.5 T (4.2 K) and do not reach the FCC target $J_c$, which serves as the present stretch target for Nb$_3$Sn development. As recently demonstrated, to meet this goal requires enhanced vortex pinning but an independent and supplementary approach is to significantly enhance $H_{c2}$. In this study, we have arc-melted multiple Nb alloys with added Hf, Zr, Ta and Ti and drawn them successfully into monofilament wires to investigate the possibilities of $H_{c2}$ enhancement through alloying. $H_{c2}$(T) was measured for all samples in fields up to 16 T and some up to 31 T. We have found that all alloys show good agreement with the standard Werthamer, Helfand, and Hohenberg (WHH) fitting procedure without the need to adjust the paramagnetic limitation parameter (α) and spin-orbit scattering parameter (λ$_{so}$). The evaluation of $dH_{c2}/dT$ near $T_c$, which is proportional to the electronic specific heat coefficient γ and the normal state resistivity ρ$_n$, allows a better understanding of the induced disorder introduced by alloying in the A15 phase. So far, we have observed that Hf alloying of pure Nb can enhance $H_{c2}(0)$ by 3-4 T to ~28 T, while adding just 1 at. %Hf or Zr into a Nb4Ta base alloy can raise $H_{c2}(0)$ to ~31 T. Very importantly we find that Hf and Zr raise the alloy recrystallization temperature above the usual A15 reaction temperature range of 650°C – 750°C, thus ensuring denser A15 phase nucleation in the Nb alloy grain boundaries, possibly leading to a more homogeneous A15 phase Sn content and refined A15 grain size. The potential for further advancements in Nb$_3$Sn properties is explored in relation to the recrystallization of the Nb alloy and the factors controlling the upper critical field.

## I. INTRODUCTION

At present, most large superconducting magnet applications like the Large Hadron Collider (LHC) at CERN and Magnetic Resonance Imaging (MRI) systems use Nb-Ti, whose practical field limit is about 12 T. Another low temperature superconductor (LTS), long used for higher field laboratory solenoid magnets, is Nb$_3$Sn. It is now being applied to accelerator dipoles, but such magnets demand the highest possible current density because their magnetic efficiency is approximately half that of solenoid magnets [1], a real challenge now that future accelerator goals for dipole magnets are well beyond 10 T (12-16 T). A recent four-layer Nb$_3$Sn dipole magnet achieved 14 T [2] but getting to a 16 T dipole is quite impossible even with the latest generation of Internal-Tin (IT) RRP (Rod Re-stacked Process) wires manufactured by Bruker Energy & SuperconTechnologies Inc. (BEST) [3]. Nuclear magnetic resonance (NMR) requires solenoid magnets up to 23 T that require sub-2 K operation for maximum field. Significant efforts have been made to optimize the non-Cu critical current density ($J_c$) which is a key parameter for such Nb$_3$Sn applications [4–6], reaching $J_c$ (4.2 K, 16 T) values up to ~ 1,375 A/mm$^2$, still below the FCC target $J_c$ of 1,500 A/mm$^2$ (4.2 K, 16 T). In Nb$_3$Sn conductors, there exists a continuous trade-off between maximizing the upper critical field ($H_{c2}$) and vortex pinning, which decides the critical current density ($J_c$) [7]. In Nb$_3$Sn conductors, the reaction temperature increase results in larger grain sizes that lower vortex pinning and current density ($J_c$) [8], but it does increase $H_{c2}$, reduce the transition breadth and raise the irreversibility field $H_{irr}$ [9], which is beneficial for practical applications. To utilize Nb$_3$Sn in the HiLumi Large Hadron Collider (LHC) upgrade, striking a good balance between maximizing the upper critical field and maximizing vortex pinning was necessary. To attain the best $J_c$(16 T) for the Future Circular Collider (FCC) will require a multifaceted improvement that will include maximizing $H_{c2}$, compositional uniformity, and enhancing GB pinning and/or "artificial pinning" (*i.e.* by oxide particles). Independently, the highest possible conversion to A15 phase formation from the so-called non-Cu package is needed too: the Nb alloy, the Sn source, any oxygen source for APC, and the diffusion barrier (DB) needed to protect the residual resistance ratio (RRR) of the stabilizing and protective copper of the conductor [6,7] must be optimized so that the highest possible cross-sectional fraction of high $J_c$ A15 is formed. In short, the further development of Nb$_3$Sn conductor technology is a complex optimization of often conflicting variables.

---

[*] Corresponding author: larbalestier@asc.magnet.fsu.edu

An underlying component of achieving the FCC goal is by maximizing $H_{c2}$. Orlando *et al.* observed that moderately increasing $\rho_n$ in Sn-deficient, binary Nb-Sn films raised $H_{c2}(0)$ to ~ 30 T, while also degrading $T_c$ to below 17 K, in strong contrast to $H_{c2}(0)$~ 21 T for clean-limit, stoichiometric Nb$_3$Sn. Their study showed that "dirty", i.e. disordered, samples could possess higher $H_{c2}$ even though being off-stoichiometric [10]. The inevitability of some Sn (and thus superconducting property) gradients in the A15 Nb$_3$Sn layer of present conductors results from a Sn activity drop as the reaction nears completion. To optimize the amount of A15 in the non-Cu package the diffusion reaction of Sn into Nb alloy to form doped Nb$_3$Sn effectively becomes an exhaustion reaction with low-Sn A15 compositions developing at the A15 reaction front during later stages of the reaction. An extensive study of these trade-offs for many reactions of binary and Nb4at.%Ta high-tin Powder-In-Tube (PIT) conductor was published by Godeke, Fischer *et al.* [9,11], which showed the way in which the compositional gradients produced by restricted reactions causes a marked lowering of the irreversibility field ($H_{irr}$) (defined by the bottom of the resistive field transition) without much affecting the top of the superconducting transition which defines $H_{c2}$ [9]. This transition breadth is tied to the Sn gradient across the A15 layer. Earlier experiments showed that $H_{c2}(0)$ varied almost linearly ~ 5 T per at% between 19.5 – 24 at% Sn while the critical transition temperature $T_c$ [12] also exhibited a linear dependence throughout the A15 phase (from 6 K at 18 at.%Sn to 18 K at stoichiometry) [13]. Beyond 24 at% Sn, $T_c$ and $H_{c2}$ no longer follow a linear composition dependence due to the cubic-tetragonal phase transformation of the A15 lattice (lattice softening produces a lower phonon frequency) [14]. The presently used Ta- and Ti-doped high $J_c$ Nb$_3$Sn conductors have $T_c$, $H_{c2}$ values that appear to show a linear dependence on at.%Sn over the whole composition range [15] because the dopants suppress the martensitic transformation, shorten the electron mean free path and raise $H_{c2}$. To better understand these alloying effects, Suenaga *et al.* [16] doped Nb with Ta, Ti, Hf, Zr and demonstrated that $H_{c2}$ was raised by ~3 T for additions of 4 at% Ta and 0.8 at% Ti. The alloying dependence of the residual resistivity shows that Ti is at least twice as effective as an A15 disordering agent as Ta [16], the increase in $H_{c2}$ correlating well to increased normal state resistivity. In contrast to Ti and Ta, Zr by itself lowered $T_c$ [17]. Later, both Ta and Ti emerged into PIT and RRP wire technology and Ti-doping by insertion of occasional Nb47Ti filaments into RRP conductors, which then became the standard method of doping RRP conductors. Heald *et al.* and Tarantini *et al.* made an extended X-ray Absorption Fine Structure (EXAFS) study on Ta, Ti, Ta + Ti RRP standard wires and showed that Ta substitutes on both Nb and Sn sites, whereas Ti substitutes only on Nb sites [18,19]. The implication is that Ti, sitting only on Nb A15 lattice sites, is a more effective disordering agent than Ta sitting on both Sn and Nb sites. Commercially, Ta, Ti, and combined Ta + Ti dopants have been explored to improve the upper critical field $H_{c2}$, but a complication is that even the best state-of-the-art modern conductors show compositional gradients across the A15 layer that cause a separation between the onset of dissipation at the irreversibility field $H_{irr}$ (where $J_c$ becomes zero) and the full restoration of the normal state at $H_{c2}$ [15]. Longer and higher temperatures, thus more complete, reactions move $H_{irr}$ closer to $H_{c2}$, which tends to be more insensitive to A15 reaction conditions [9]. We must therefore be concerned more about $H_{irr}$ than $H_{c2}$ for conductor optimization [9,19]. Complications due to strain either due to the tetragonal phase transition or operational stresses are considered to be second order effects when the conductor architecture is kept constant, and minimization of strain is practiced as is the case in all the experiments reported here.

Time of reaction may be of interest too for magnet builders and various studies have shown that different alloying element additions have effects on the A15 reaction rates, Ti for example producing faster reaction than Ta additions [20,21]. It has also been observed that Zn, Al, or Mg additions increase the rate of Nb$_3$Sn formation, even if they do not dissolve into the Nb$_3$Sn layer [22]. However, Ta, Hf, and Ga do dissolve into the A15 layer and thus can influence the intrinsic superconducting properties [17,22–25]. For instance, Ga and Hf additions to Nb$_3$Sn have been reported to improve $J_c$(16 T) by one order of magnitude, $T_c$ by 1 K and upper critical field by 5 T with respect to binary Nb$_3$Sn [26–28], though few studies showed good quality high field $H_{c2}$ transitions, making it uncertain for a long time whether

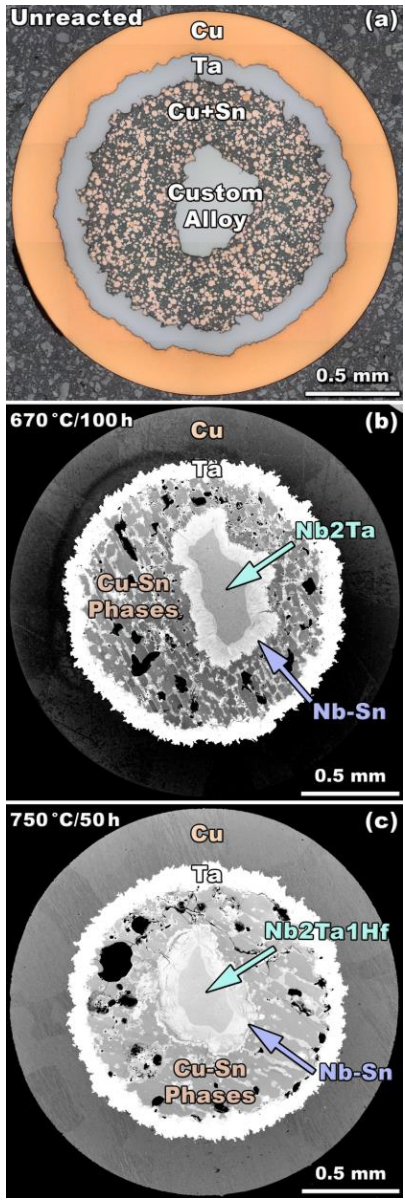

FIG. 1 (a) Light microscope image of a cross-section of an unreacted monofilament wire. In-house melted custom alloy rods were placed in the center of a Cu+Sn powder mixture fitted inside a commercial seam welded Ta-tube which was sheathed in a Cu-tube. (b) and (c) Typical FESEM-BSE cross-sections of reacted monofilament wires. The light contrast at the perimeter of the central alloy rods reveals the reacted layers.

studies of new alloys were justified or not. Only with the issuance of the FCC specification has the issue become so important that the optimum alloying of $Nb_3Sn$ deserves revisiting [29–32]. The precursor to this study was recent work in ASC-NHMFL and some other groups which shows that the addition of group IV (Hf, Zr) elements to Nb4at%Ta enhances $H_{irr}$, although the addition of $O_2$ might slightly degrade it [29,30,32,33]. The best present state-of-the-art Ta- and Ti-doped RRP conductors show $H_{c2}$(4.2 K) values in the range 24 – 26.5 T. But, as noted above, the overall goal must always be to develop a high field conductor with the finest possible A15 grain size and minimum $H_{irr} - H_{c2}$ transition breadth. To this end, this paper focuses on the following issues:

1) Developing a thorough understanding of how the upper critical field of $Nb_3Sn$ evolves as it is doped with elements such as Ta, Hf, Zr, Ti and Ga that enter the A15 phase. We used our arc-melter to make Nb alloys of many compositions and then reacted them by diffusion reaction with Cu-Sn and Cu-Ga-Sn mixtures.
2) Sekine *et al.* reported that additions of Hf, and Ga and Hf together significantly enhanced the $H_{c2}$ of $Nb_3Sn$ conductors [27]. Accordingly, we restudied the effect of adding Hf to Nb (with Ga added to the Cu-Sn mixture), but also investigated the Hf-Ga effectiveness when combined with the NbTa alloy, which by itself leads to a higher $Nb_3Sn$ $H_{c2}$ with respect to the use of pure Nb. This more complex alloy (NbTaHf) is in fact a candidate for the realization of new high field conductors.
3) With present Ti and Ta-doped Nb, $H_{c2}(0)$ can be increased from 25–27 T to less than 30 T [9,15,18]. Because these enhancements are relatively small, we made careful measurements of $H_{c2}(T)$ for all our alloys and investigated whether the $H_{c2}$ enhancement requires consideration of paramagnetic limiting or spin-orbit scattering effects, finding that they do not. We show that our optimum addition of both Ta

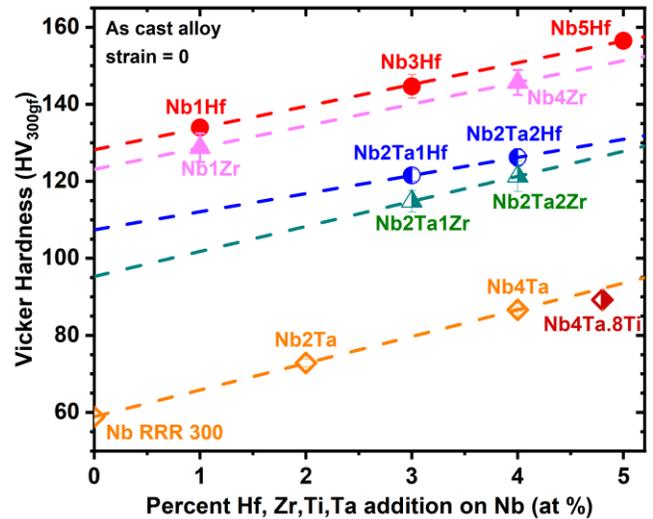

FIG. 2 Vickers hardness as a function of EDS-determined alloy compositions. The Vickers hardness increases in all cases as the dopant addition increases.

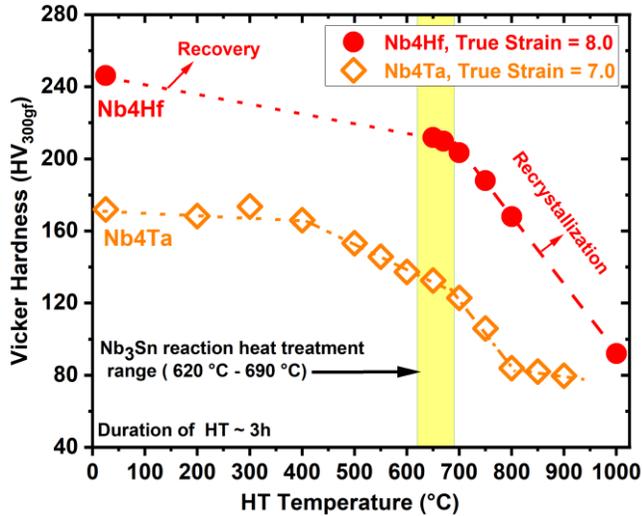

FIG. 3 Vicker hardness after 3 hour anneals at the stated heat treatment (HT) temperatures used to study the recrystallization behavior of Nb4Ta and Nb4Hf. It is clear that Hf significantly delays recrystallization as compared to the equivalent Nb-Ta alloy.

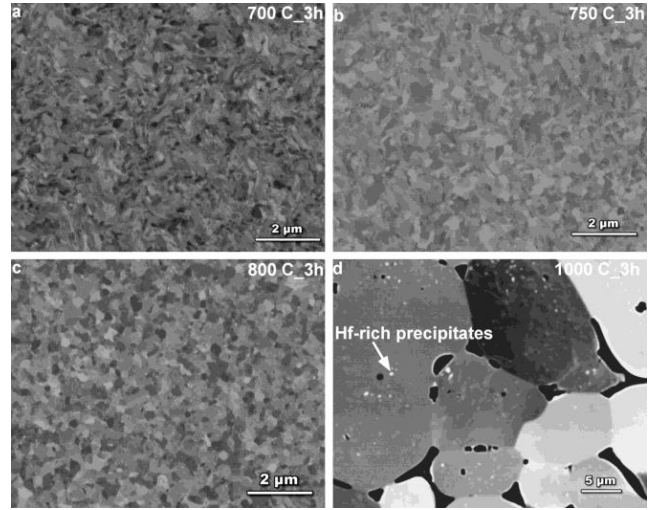

FIG. 4 Microstructural evolution of Nb4Hf from (a) 700 °C for 3 h, (b) 750 °C for 3 h, (c) 800 °C for 3 h, and (d) 1000 °C for 3 h (FESEM-BSE images). The annealing of Nb4Hf have little effect on grain size even up to 800 °C and grain growth is only visible at an annealing temperature of 1000 °C.

and Hf increases $H_{c2}(0)$ to over 30 T.

4) Because A15 grain refinement is critical to enhancing vortex pinning, we studied the effect of alloying on the stability of the cold-worked alloy grain structures, finding that both Zr and Hf are effective in inhibiting recovery and recrystallization of the Nb-alloy to above the normal A15 reaction temperature range [30,32,33].

## II. EXPERIMENTAL METHODS

We arc-melted custom alloys by using commercial Nb$x$Ta ($x$=0,2,4at.%) base alloys and adding Hf, Zr, Ti by themselves or in combination. All Nb alloys are referred to by the commonly used notation specifying the at% of their dopant(s) (e.g. Nb2Ta2Hf is the alloy made with 96 at% Nb, 2 at% Ta and 2 at% Hf) unless otherwise specified. Our in-house, arc-melter was used to produce batches of ~35 g ingots. Loss of material during the melting process was negligible (< 0.2 g).

Vickers hardness tests, SQUID magnetometer characterizations, and SEM analyses were conducted on the as-cast Nb-alloys to evaluate their quality before they were drawn into wires. All alloys had a high $T_c$ onset (≥ 9 K) and sharp transitions (≤ 0.25 K). Our wire design, consisted in surrounding a Nb-alloy core rod with Cu+Sn(+Ga) powder, and then further cladding with a composite Ta/Cu tube as in [32], as depicted in Figure 1. The central alloy rod was swaged from the irregular arc-melted billet to a 3 mm diameter, 400 mm long rod. The Cu-Sn powders used had an initial mesh size of 320 (~50 μm) and were mixed in a 6:5 atomic ratio. The initial package was drawn to a final wire diameter of 2 mm. The total true strain (ε) is determined by the cross-sectional area reduction, ε= ln(A$_0$/A$_f$), where A$_0$ is the initial area, and A$_f$ is the final area.

To determine the Vickers hardness of the as-cast alloy and its correlation with the true drawing strain, all arc-melted ingots were swaged using a 20% reduction in area and samples were taken every 2-3 passes. The Vickers hardness was measured on metallographically polished transverse sections using a 300 g load. The SQUID magnetometer measurements from 4.0 K − 12.0 K were taken on the alloys showing transition width of typically less than 0.3 K. The monofilament testbeds were made with Cu+Sn and Cu+Sn+Ga powders. The starting composition of Cu+Sn powder was Cu45at%Sn (Cu$_6$Sn$_5$ equivalent), whereas the used Cu+Sn+Ga powder mixes were Cu25at%Sn20at%Ga and Cu30at%Sn2at%Ga. The monofilament testbeds were drawn to a filament diameter of 2 mm followed by a two-step heat treatment. A 550 °C hold was previously chosen to decompose SnO$_2$ in wires containing it, however it is also used in this paper for direct comparison to the results obtained earlier on Hf/Zr/Ta doped samples using the same heat treatment schedule [32,34]. A second step ranging between 650 °C and 850 °C was used for the A15 reactions. Prior to transport measurements the external Cu was etched away, and the Ta tube cut open to extract the central alloyed-Nb rod surrounded by the A15 layer on which the contacts were made. Resistive transport characterizations to determine $H_{c2}$(T) were performed in our 16 T Quantum

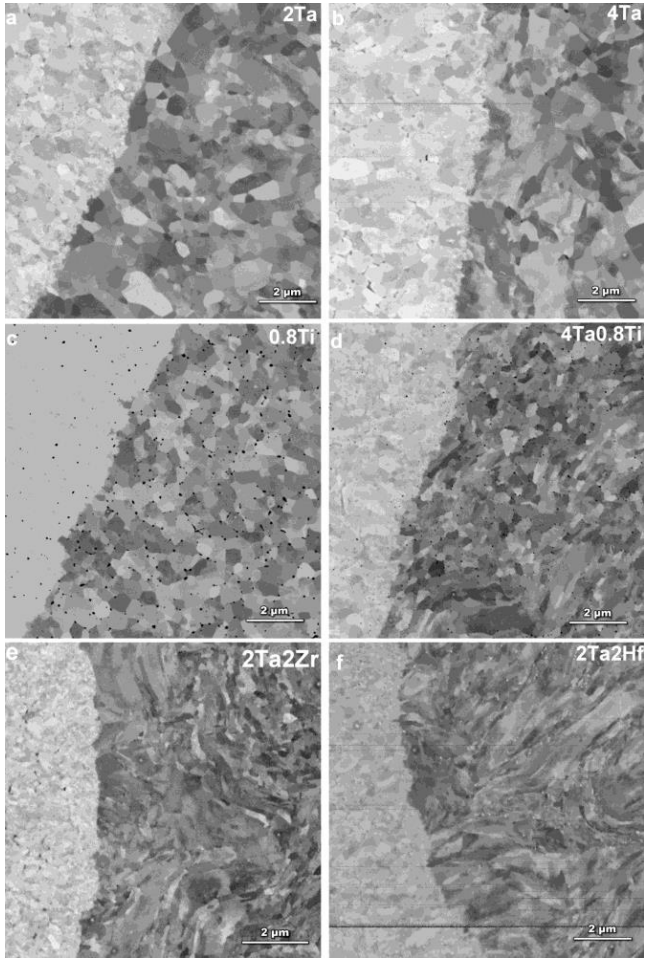

FIG. 5 Representative FESEM-BSE images of the Nb-alloy core/Nb$_3$Sn interface after the monofilaments were reacted at 550 °C/50h + 750 °C/50 h. The darker Nb-alloy cores are on the right and the lighter A15 diffusion layers are on the left of the interface (sample IDs are indicated in the top right corner of each image). Electron channeling contrast in the corresponding images reveals recrystallized microstructures in the unreacted Nb$x$Ta($x$=2,4), Nb0.8Ti, and Nb4Ta0.8Ti alloys. In contrast, the Nb2Ta2Hf and Nb2Ta2Zr alloys retain a deformed, cold-worked grain structure.

Design PPMS and some wires were measured up to 31 T at the National High Magnetic Field Laboratory (NHMFL). A Zeiss 1540 EsB field-emission scanning electron microscope (FESEM) was used to evaluate the microstructures on transverse wire cross-sections prepared by metallographic polishing. The samples were imaged with a solid-state back-scattered electron (BSE) detector, and energy dispersive X-ray spectroscopy (EDS) was used to study the elemental compositions of the as-cast alloys and their A15 layers. In the following we will identify the doped Nb$_3$Sn samples according to their at.% doping content only (e.g. 4Ta1Hf is the doped Nb$_3$Sn wire made with the Nb4Ta1Hf alloy).

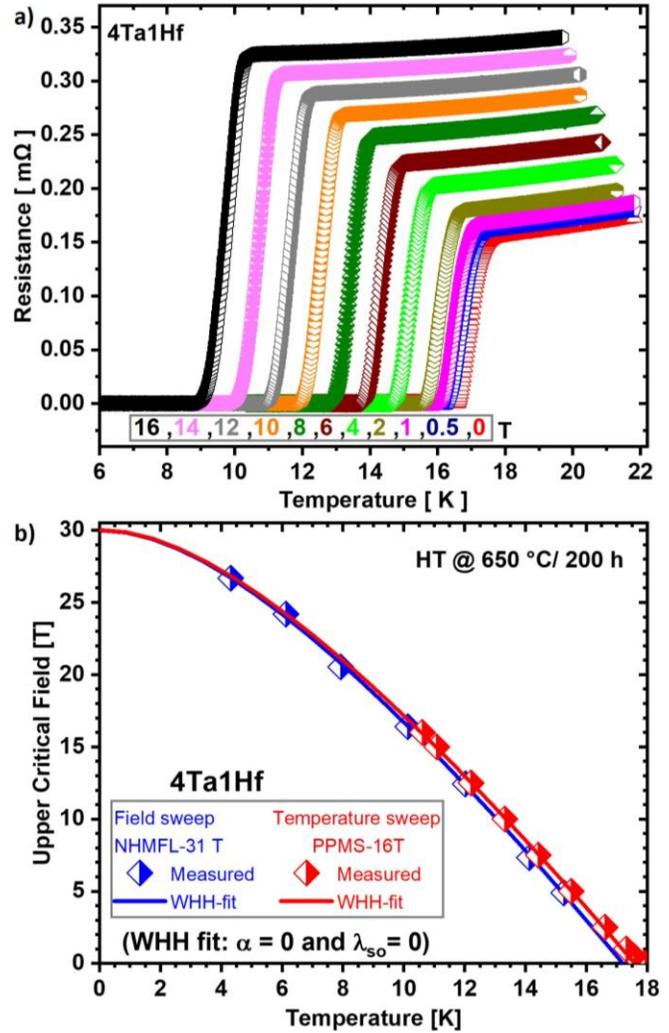

FIG. 6 (a) The traces of resistance against temperature using temperature sweeps at constant field from 0 – 16 T in our 16 T PPMS for Nb4Ta1Hf reacted at 550 °C/50h + 650 °C/200 h; (b) At bottom we show the experimental $H_{c2}$(T) values and their WHH fits in PPMS-16T and NHMFL 31T.

## III. RESULTS

### A. Recrystallization and microstructural characterizations

The Vickers hardness H$_V$ was used to evaluate the potential wire fabricability and its response to heat treatment in the A15 reaction window. H$_v$ versus alloy content and type is shown in Figure 2. All alloys in both the as-cast and worked states were hardened by additions of Ta, Zr and Hf. Consistent with their size difference, the larger Hf and Zr atoms hardened the Nb more strongly than Ta, but this did not make the alloys challenging to draw. The increase in hardness with increasing Hf, Zr and Ta solute in the binary/ternary alloys is linear with the total percentage of addition in all cases. A comparison of the recrystallization behaviour of Nb4Hf and Nb4Ta after 3 hour anneals at different temperatures is presented in Figure 3, which clearly shows that $H_V$ starts to decline above about 400 °C for Nb4Ta but also that this is delayed to 650-700 °C for the Nb4Hf alloy. The initial recovery processes occurring within the grains shown in Figure 4 have little effect on grain size even up to 800 °C and real grain growth for the Nb4Hf alloy is only visible in the sample annealed at 1000 °C, which is consistent with the hardness change behaviour shown in Figure 3. We also see some high atomic number (light) precipitates in the recrystallized structure (Fig. 4(d)), which the SEM-EDS analysis shows to be Hf-rich; those are more evident after the high-temperature heat treatment. FESEM-BSE imaging revealed both the $Nb_3Sn$ A15 layer and the partly annealed alloy structure in the unreacted cores in Figure 5. The evident recrystallization of the Nb2Ta, Nb4Ta, Nb0.8Ti, and Nb4Ta0.8Ti alloy cores is highlighted by channelling contrast in the BSE images. In contrast, the Hf and Zr alloyed (Nb2Ta2Hf, Nb2Ta2Zr) core rods retain their worked structure, even after A15 reactions at 550 °C/50h + 750 °C/50 h.

## B. $H_{c2}$(T) and its analysis

Small current (~10 mA), resistive transport characterizations, R(T,H), to determine $H_{c2}$(T) were performed with contacts on the bare $Nb_3Sn$ layers on all the reacted wires in the 16 T PPMS using temperature sweeps over the range 1.9 K- 22 K. Some wires were also measured in higher fields in the 31 T resistive magnet at the NHMFL using constant-temperature field sweeps in the range 0 – 31 T. Typical resistive transition traces using temperature sweeps at constant field are shown in Figure 6a for a 4Ta1Hf wire (the normal state magnetoresistance originates in the residual unreacted Nb alloy, not the A15 layer). $H_{c2}$(T) was evaluated at 90% of the normal state resistance for both PPMS-16 T and 31 T data sets (two pieces of the same wire were used). The data in Figure 6b is fitted by the full WHH relation [35]:

$$ln\frac{1}{t} = \left(\frac{1}{2} + \frac{i\lambda_{so}}{4\gamma}\right)\psi\left(\frac{1}{2} + \frac{\bar{h} + \frac{1}{2}\lambda_{so} + i\gamma}{2t}\right) +$$

$$+ \left(\frac{1}{2} - \frac{i\lambda_{so}}{4\gamma}\right)\psi\left(\frac{1}{2} + \frac{\bar{h} + \frac{1}{2}\lambda_{so} - i\gamma}{2t}\right) - \psi\left(\frac{1}{2}\right)$$

(where $\gamma = \left[(\alpha\bar{h})^2 - \left(\frac{1}{2}\lambda_{so}\right)^2\right]$ , $t = \frac{T}{T_c}$ , $\bar{h} = \frac{4\,H_{c2}(T)}{\pi^2 T_c \left|\frac{dH_{c2}}{dT}\right|_{T_c}}$ and $\psi$ being the digamma functions) and they are well reproduced with the paramagnetic limitation parameter ($\alpha$) and a spin-orbit scattering parameter $\lambda_{so}$) both set to 0. This implies that the $Nb_3Sn$ data can be fitted with the reduced $ln\frac{1}{t} = \psi\left(\frac{1}{2} + \frac{\bar{h}}{2t}\right) - \psi\left(\frac{1}{2}\right)$ relation. In the following data discussions, the parameters derived from the WHH fits are marked with "WHH" superscript (e.g. $T_c^{WHH}$). The WHH theory predicts that for a=0 and $\lambda_{so}$ =0, the 0 K limit is given by the relation $H_{c2}(0) \sim 0.693\, T_c \left|\frac{dH_{c2}}{dT}\right|_{T_c}$. This simpler expression instead of the full WHH fit has been widely used to estimate $H_{c2}(0)$ values of many superconductors since all it requires is to measure the $H_{c2}$ slope in the higher-temperature, lower-field limit. There are, however, some experimental concerns when numerically accurate predictions are needed. Many samples have compositional variations for various reasons, as is the present case, because Nb$_3$Sn is not a line compound and, when made by diffusion, as here, composition gradients are always present. Because of the different dependences of $T_c$ and $H_{c2}$ on composition, this may lead to non-WHH $H_{c2}$ behaviour near $T_c$. Since we have to take this into account to analyse our data, we explain in detail what causes these possible anomalies and the method we used to analyse our $H_{c2}$ data within the WHH framework.

TABLE I. A15 superconducting parameters for samples reacted at 550 °C/100h + 670 °C/100h and 550 °C/50h + 750 °C/50h (experimental and fitting parameter errors are typically defined by the last indicated digit).

| Sample ID | $T_{c,exp.}$ [K] | | $T_c^{WHH}$ [K] | | $\mu_0 H_{c2}^{WHH}(0)$ [T] | | $\mu_0 \left|\frac{dH_{c2}^{WHH}(0)}{dT}\right|_{T_c}$ $\left[\frac{T}{K}\right]$ | |
|---|---|---|---|---|---|---|---|---|
| | 670 °C / 100h | 750 °C / 50h | 670 °C / 100h | 750 °C / 50h | 670 °C / 100h | 750 °C / 50h | 670 °C / 100h | 750 °C / 50h |
| 4Ta | 17.31 | 17.72 | 17.24 | 17.40 | 29.0 | 28.9 | 2.43 | 2.40 |
| 2Ta | 17.61 | 18.15 | 17.49 | 17.86 | 28.2 | 28.5 | 2.33 | 2.31 |
| 4Ta1Hf | 17.18 | 17.77 | 17.19 | 17.59 | 30.2 | 29.2 | 2.54 | 2.40 |
| 4Ta2Hf | 17.57 | ---- | 17.46 | ---- | 29.7 | ---- | 2.46 | ---- |
| 2Ta1Hf | 16.96 | 17.87 | 16.99 | 17.69 | 28.3 | 28.8 | 2.40 | 2.35 |
| 2Ta2Hf | 17.32 | 18.08 | 17.30 | 17.81 | 29.4 | 28.9 | 2.45 | 2.34 |
| 4Ta1Zr | 17.44 | ---- | 17.36 | ---- | 30.0 | ---- | 2.49 | ---- |
| 2Ta1Zr | 16.99 | 17.72 | 16.98 | 17.53 | 27.9 | 27.9 | 2.37 | 2.30 |
| 2Ta2Zr | 17.35 | 17.85 | 17.35 | 17.69 | 29.3 | 28.7 | 2.44 | 2.34 |
| 2Ta0.5Hf0.5Zr | 17.24 | 17.81 | 17.23 | 17.62 | 29.2 | 28.7 | 2.45 | 2.35 |
| 2Ta1Hf1Zr | 17.20 | 17.94 | 17.27 | 17.73 | 29.0 | 28.6 | 2.42 | 2.33 |
| 1Hf | 17.17 | 17.75 | 17.11 | 17.52 | 27.9 | 27.2 | 2.35 | 2.24 |
| 3Hf | 17.38 | 18.03 | 17.34 | 17.66 | 27.4 | 26.9 | 2.28 | 2.20 |
| 5Hf | 17.49 | 17.85 | 17.41 | 17.51 | 28.0 | 26.1 | 2.32 | 2.15 |
| 0.8Ti | 17.67 | ---- | 17.52 | ---- | 26.8 | ---- | 2.21 | ---- |
| 0.8Ti4Ta | 17.34 | ---- | 17.22 | ---- | 27.7 | ---- | 2.32 | ---- |
| 0.8Ti1Hf | 17.10 | ---- | 17.03 | ---- | 26.8 | ---- | 2.27 | ----- |

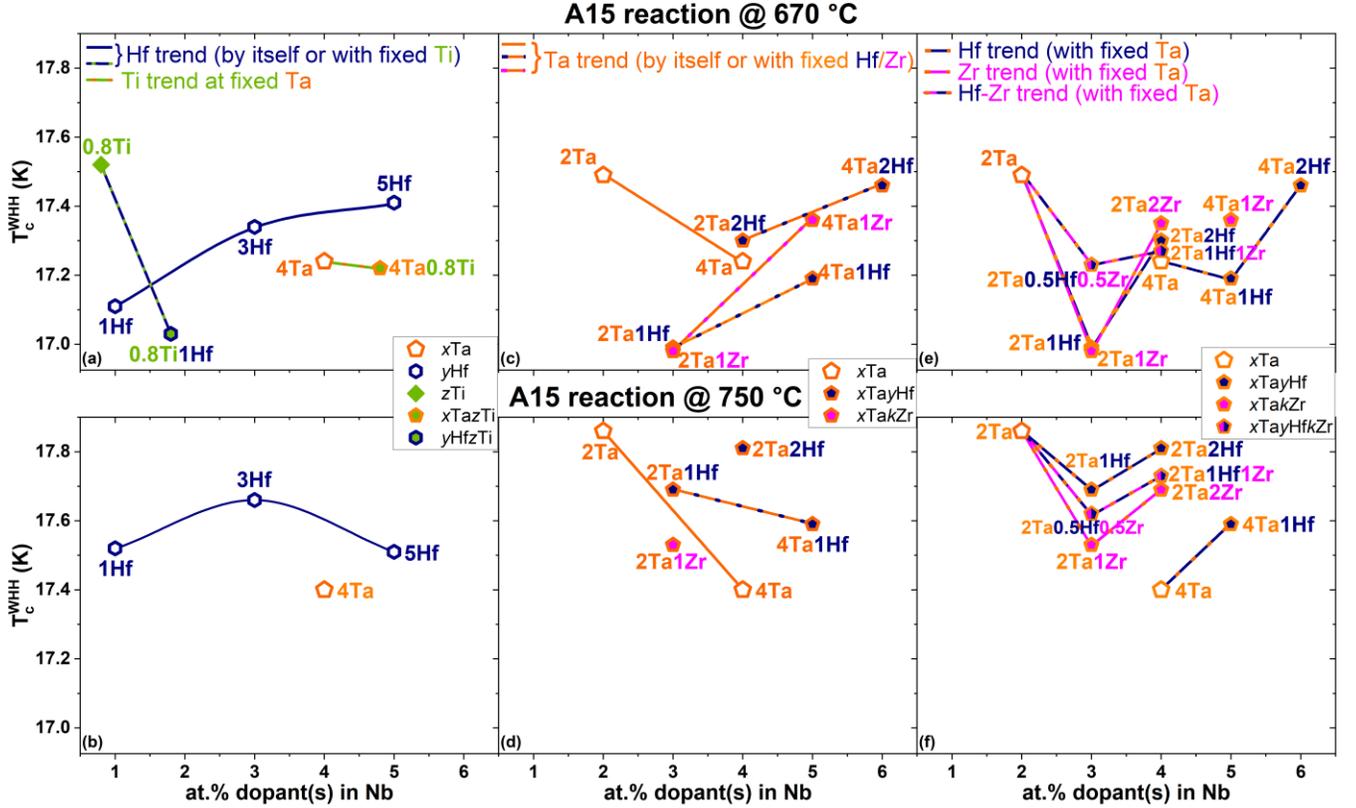

FIG. 7 $T_c$ evaluated by the WHH fits for alloyed A15 samples reacted at 550 °C/100 h + 670 °C/100 h (top row) and at 550 °C/50 h + 750 °C/50 h (bottom row) as a function of the amount of dopant(s). For clarity the data has been organized into 3 columns (a-b, c-d, and e-f) of alloy sets with relevant fixed versus variable element concentrations. As explained in the text the labels refer to the total at.% of dopant(s) in the unreacted Nb alloys. It is immediately obvious that the trends are not the same at the two reaction temperatures.

In inhomogeneous, varying composition samples both a $T_c$ and an $H_{c2}$ distribution result. Since typically the sample fraction with the highest $T_c$ is closer to stoichiometry, it does not necessarily produce the highest $H_{c2}$ at the lowest temperature. The result is thus a crossover in the $H_{c2}(T)$ trends of different portions of the A15 layer, which may produce a mild upward curvature near $T_c$, instead of a completely linear $H_{c2}(T)$. An $H_{c2}$ slope determined in this upturn region would inevitably underestimate the $H_{c2}$ slope of the sample fraction that generates the highest low-temperature $H_{c2}$, so we did not use this approach. Moreover, properly evaluating the $H_{c2}$ slope near $T_c$ is also difficult for homogeneous sample. In fact, according to the WHH relation, a linear fit of a range above $t=0.9$ causes an underestimation of the slope of ~2%, which increases at ~4% if fitted above $t=0.8$. For these reasons, to exclude a possible effect of an upward curvature near $T_c$, we prefer to analyse our $H_{c2}(T)$ results with the WHH fits but only on data at $\mu_0 H \geq 2$ T. The correctness of this approach is confirmed by both the quality of the fits and the closeness of the fitting parameter $T_c^{WHH}$ to the experimentally evaluated $T_{c,\exp}$, reported in Table 1. As expected, $T_c^{WHH}$ is within the experimental error of $T_{c,\exp}$ or only slightly smaller, indicating minor inhomogeneity in only a few samples. As presented below, from the WHH fits we determine the essential parameters $T_c^{WHH}$ and $|dH_{c2}^{WHH}/dT|_{T_c}$ (or equivalently $H_{c2}^{WHH}(0)$). The value of determining and comparing the $H_{c2}$ slopes, which is a convolution of $\gamma$ and $\rho_n$, will be discussed in section 4.

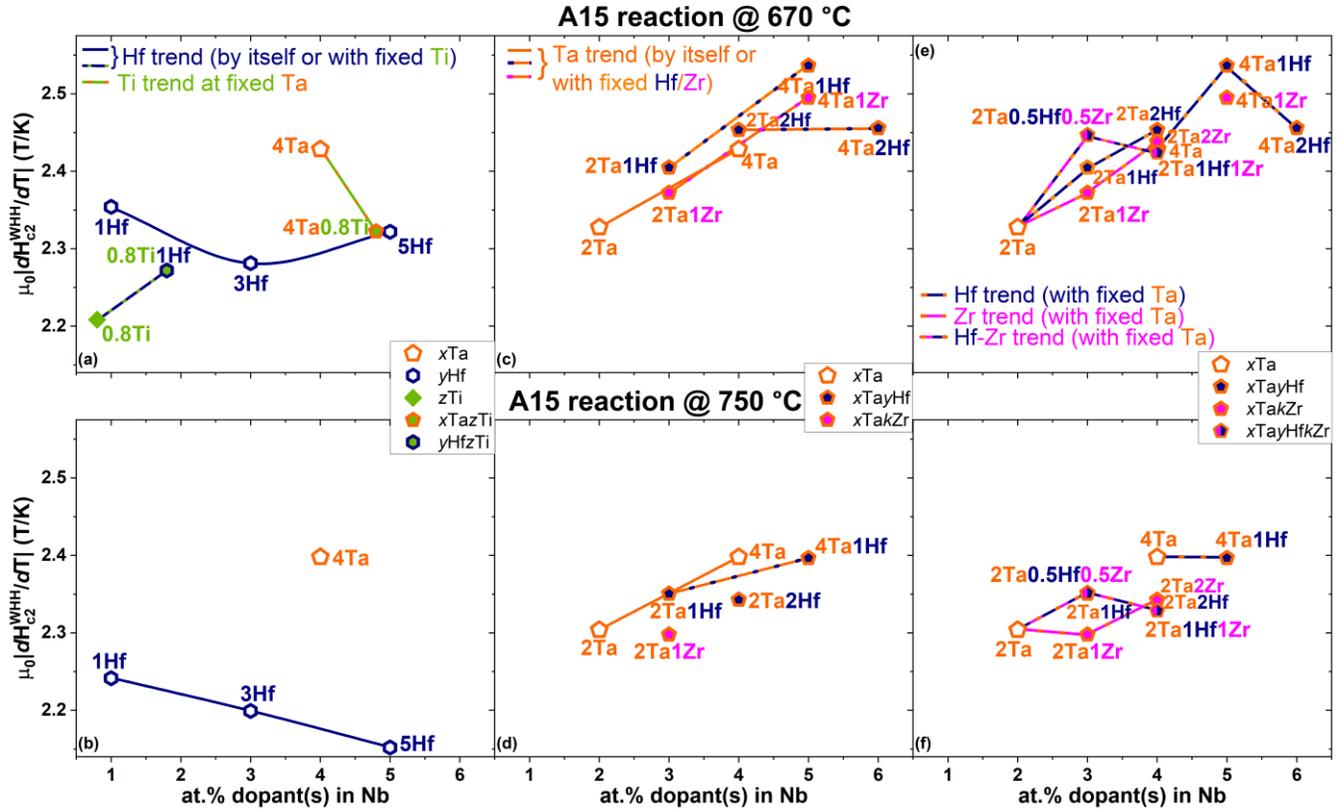

FIG. 8 $H_{c2}$ slope evaluated by the WHH fits for alloyed A15 samples reacted at 550 °C/100 h + 670 °C/100 h (top row) and at 550 °C/50 h + 750 °C/50 h (bottom row) as a function of the amount of dopant(s). The same alloy sets as used for the a-b, c-d and d-e columns in Fig. 7 are used for the column sets here. As explained in the text the labels refer to the at.% of dopant(s) in the unreacted Nb alloys.

### *1. Effects of alloy dopants*

To better understand how the dopants affect the different properties, we plotted in Figure 7 the $T_c$ data as a function of the total percentage of dopants in the unreacted alloys and for clarity we separated the different trends into multiple panels: the data in the top panels are for an A15 reaction at 670 °C, whereas those in the bottom row are obtained after a 750 °C reaction. The left panels show the effect of Hf by itself or with fixed Ti, and the effect of Ti at fixed Ta. The middle panels report the effect of Ta by itself or when added to a fixed amount of Hf or Zr. The right panels represent the effect of Hf, Zr or a combination of the two when added to a fixed amount of Ta. Equivalent data for the slope of $H_{c2}^{WHH}$ and $H_{c2}^{WHH}(0)$, described in the following are reported similarly in Figures 8 and 9. The effects of dopants on $T_c^{WHH}$ are quite complex and not always monotonic, but a general feature is that $T_c^{WHH}$ increase with increasing the reaction temperature; the maximum difference (0.7 K) is observed in the 2Ta1Hf case.

Adding Hf to Ti is clearly detrimental (Fig. 7a), whereas adding Hf by itself has a peculiar effect. With only 1%Hf $T_c^{WHH}$ is clearly depressed from the typical ~18 K of the binary Nb$_3$Sn, but it recovers at higher Hf level for the 670 °C heat treatment (Fig.7a); $T_c^{WHH}$ is higher and non-monotonic at 750 °C (Fig. 7b). In our sample we also observed a decrease of $T_c^{WHH}$ with increasing the Ta content by itself from 2 to 4 % (Figs.7c-d), whereas increasing Ta in combination with a fixed amount of Hf or Zr is beneficial to $T_c^{WHH}$ or causes small suppression (Figs.7c-d); none of these double-doped wires have however a better $T_c$ of the 2Ta wire. A very interesting trend is observed when increasing the Hf, Zr or Hf-Zr content in the presence of a fixed amount of Ta (Figs.7e-f). In almost all cases (with the exception of 4Ta/4Ta1Hf at 750 °C) the trends show a V-shape: in fact, the good $T_c^{WHH}$ observed for $x$Ta is reduced by adding 1% of other dopants to Ta (i.e. 2Ta1Hf, 4Ta1Hf at 670 °C only, 2Ta1Zr or 2Ta0.5Hf05Zr), whereas adding 2% of other dopants increases $T_c^{WHH}$ again.

Increasing the reaction temperature causes a decrease of the $H_{c2}$ slope in all cases, so an opposite behaviour with respect to $T_c^{WHH}$ (Fig. 8). Hf addition to Nb0.8Ti increases the $H_{c2}$ slope, whereas the Ti addition to the Nb4Ta decreases it (Figs. 8a). Increasing the amount of Hf by itself has a small non-systematic variation on the $H_{c2}^{WHH}$ slope at 670 °C, but it is clearly detrimental at 750 °C. Significantly higher values of $|dH_{c2}^{WHH}/dT|_{T_c}$ are obtained for various combinations of Ta, Hf and/or Zr at 670 °C. Fig. 8c reveals that increasing the Ta content by itself or in the presence of Hf or Zr increases the slope: in fact, it increases from 2.33 to 2.43 T/K going from 2Ta to 4 Ta. In the presence of 1Zr, the Ta trend at 670 °C is similar to that induced by the Ta only, reaching 2.49 T/K in the 4Ta1Zr case. Better is the trend found for Ta in combination of Hf, with a slope that increases from 2.40 to 2.54 T/K going from 2Ta1Hf to 4Ta1Hf, with the latter being the highest value of all. In the presence of 2Hf, increasing Ta has no significant impact. At 750 °C (Fig. 8d), the behaviour is different and the presence of Hf or Zr has limited or no effect in increasing $|dH_{c2}^{WHH}/dT|_{T_c}$ with respect to the only Ta trend. Looking at the trends with fixed Ta amount (Fig. 8e-f), adding an increasing amount of Hf or Zr to the 2Ta-base alloy is beneficial for the $H_{c2}^{WHH}$ slope at 670 °C (with Hf being the more effective, Fig. 8e), whereas adding Hf and Zr together to the 2Ta-base has a non-monotonic behaviour (inverted V-shape) with the 2Ta0.5Hf0.5Zr (2.45 T/K) being better than 2Ta (2.33 T/K) and 2Ta1Hf1Zr (2.42 T/K). A similar non-monotonic trend is also observed for the 4TayHf series at 670 °C. Smaller and non-monotonic variations are found at 750 °C (Fig. 8f).

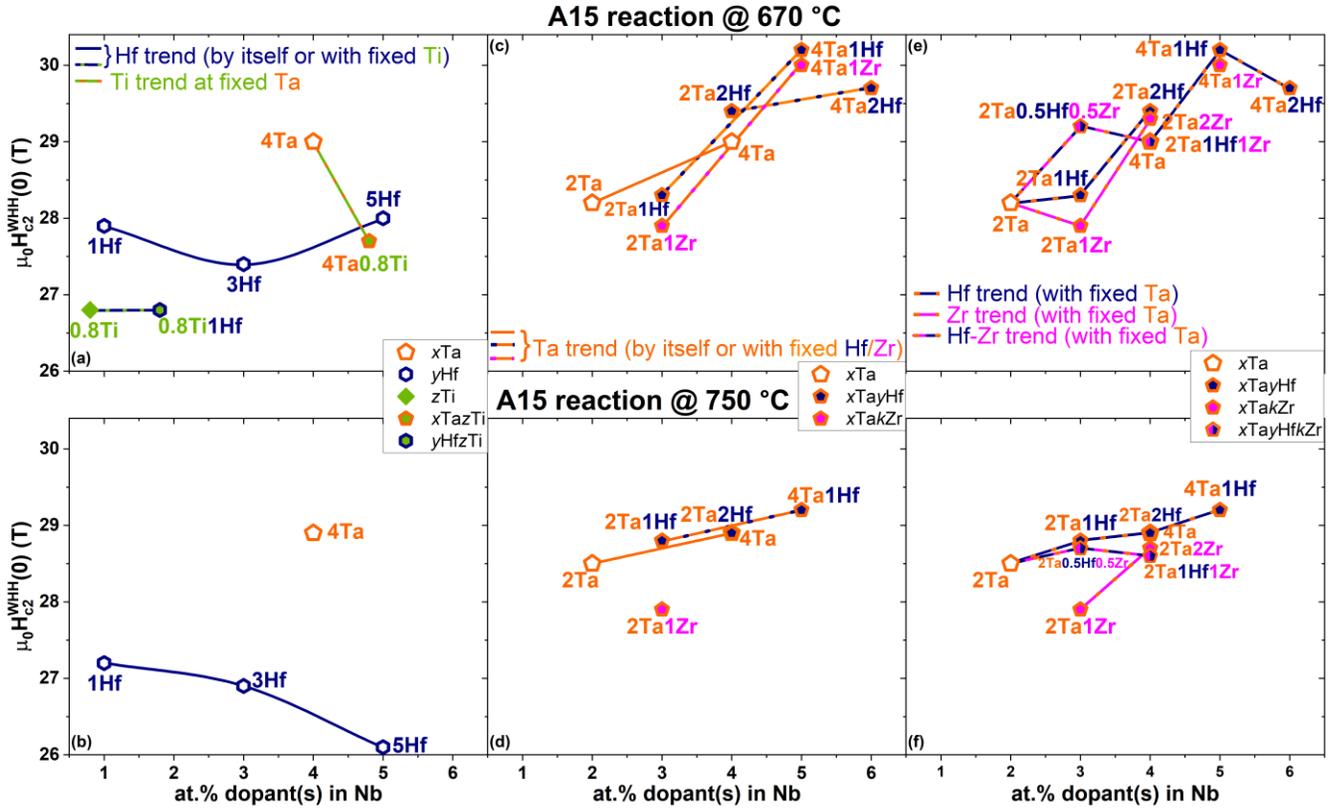

FIG. 9 $H_{c2}(0)$ evaluated by the WHH fits for alloyed A15 samples reacted at 550 °C/100 h + 670 °C/100 h (top row) and at 550 °C/50 h + 750 °C/50 h (bottom row) as a function of the amount of dopant(s). The same alloy sets as used for the a-b, c-d and d-e columns in Fig. 7 are used for the column sets here. As explained in the text the labels refer to the at.% of dopant(s) in the unreacted Nb alloys.

Since we are ultimately interested in the low temperature performance, it is important to also verify the trends in $H_{c2}^{WHH}(0)$, which is proportional to $T_c^{WHH}$ and $|dH_{c2}^{WHH}/dT|_{T_c}$. Figure 9 shows that, except for 2Ta1Hf (+0.5 T), 2Ta1Zr (no difference) and the $x$Ta series (small variations), all other samples reacted at 750 °C have a lower $H_{c2}^{WHH}(0)$ than those at 670 °C. At the lower reaction temperature (Fig.9a), Hf by itself or in combination with Ti do not generate high $H_{c2}^{WHH}(0)$ (28 T maximum) and 5Hf at 750 °C causes the strongest $H_{c2}$ suppression (26.1 T in Fig.9b). 4Ta leads to an $H_{c2}^{WHH}(0)$ of 29 T, compared to 28.2 T for 2Ta (Figs. 9c). In the presence of 1%Hf (Fig. 9c), $H_{c2}$ is enhanced much more markedly, going from 2Ta1Hf (28.3 T) to 4Ta1Hf (30.2 T); adding 1%Zr to $x$Ta produces a similar trend but with slightly lower $H_{c2}^{WHH}(0)$ than with the Hf (27.9 and 30.0 T in 2Ta1Zr and 4Ta1Zr, respectively). The $x$Ta2Hf series is clearly better than the $x$Ta one (Fig. 9c), but it does not reach the performance of 4Ta1Hf or 4Ta1Zr. At

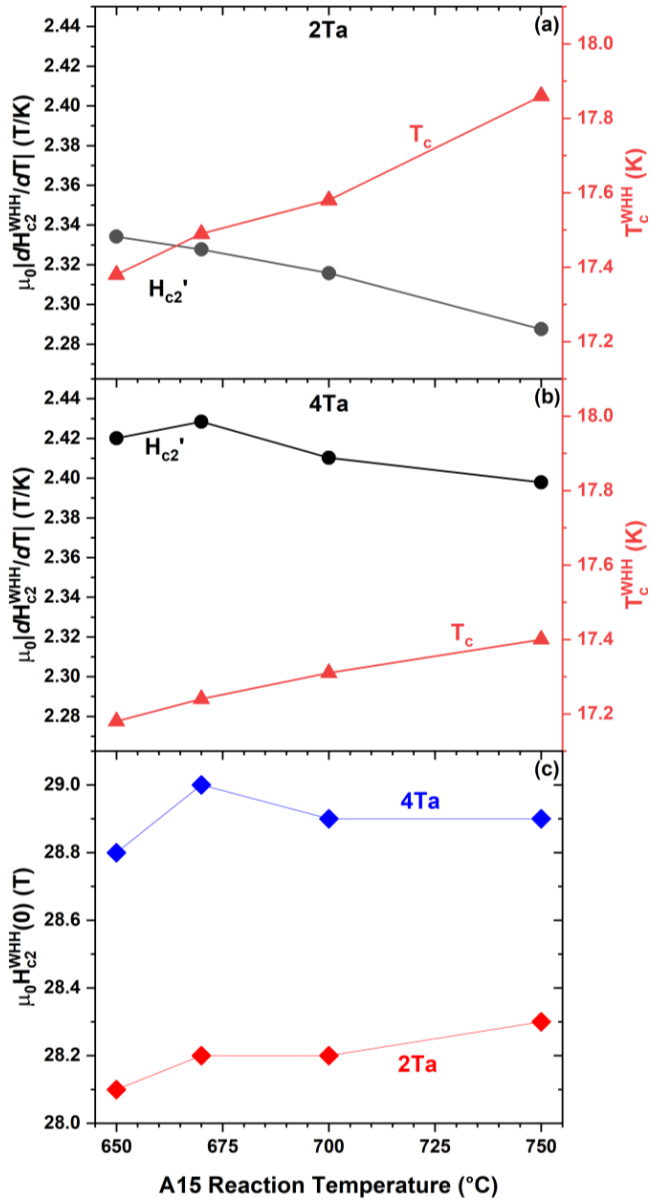

FIG. 10 Variation of $|dH_{c2}^{WHH}/dT|_{T_c}$ and $T_c^{WHH}$ in the (a) 2Ta and (b) 4Ta wires as a function of A15 reaction temperature. (c) Variation of $H_{c2}^{WHH}(0)$ for both sample series.

750 °C (Fig. 9d) the trends are closer and mostly determined by the total dopant content, with the exception on 2Ta1Zr that has a lower $H_{c2}^{WHH}(0)$ with respect to all other samples.

Changing the Hf, Zr or Hf+Zr content in the 2Ta- and 4Ta-bases (Figs. 9e-f) produced non-monotonic behaviours in all 670 °C cases, although with different trends depending on whether $T_c$ or the $H_{c2}^{WHH}$ slope dominates. The best $H_{c2}^{WHH}(0)$ is reached in 4Ta1Hf, and a further increase of Hf is not beneficial. A total of 4% of dopants (4Ta, 2Ta2Hf, 2Ta2Zr or 2Ta1Hf1Zr) produces a relatively small variation in $H_{c2}^{WHH}(0)$ in both reactions (29-29.4 T at 670 °C and 28.6-28.9 T at 750 °C). Overall, despite the spreading on the data, Figure 9 shows that (excluding Ti and single Hf additions) there is an almost monotonic increase in $H_{c2}^{WHH}(0)$ with increasing overall dopant content up to 5 at.%, whether by Ta alone or by Ta+Hf, Ta+Zr or Ta+Hf+Zr combinations.

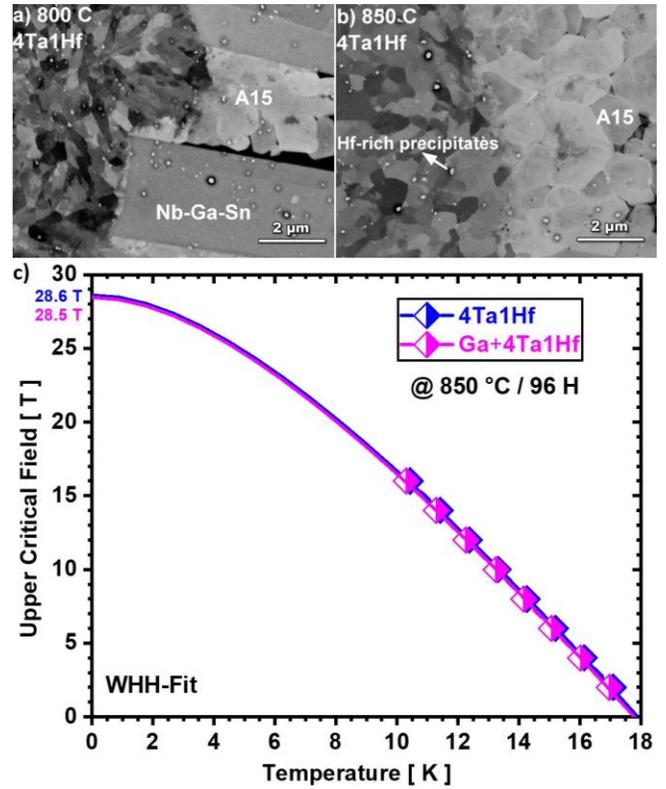

Figure 11. FESEM-BSE images of the Ga+4Ta1Hf wires made with Nb4Ta1Hf + Cu-Ga-Sn and reacted at (a) 800 °C, (b) 850 °C and (c) WHH fit of upper critical field for the and Ga+4Ta1Hf reacted at 550 °C/50h + 850 °C/96h. SEM-EDS analysis shows up to 1.7 at% Ga in A15 layer at 850 °C. The $H_{c2}(0)$ of Ga+4Ta1Hf reacted at 850 °C are 2 T lower than for 4Ta1Hf without Ga reacted at much lower temperature 670 °C.

### *2. Effects of reaction temperature on Ta-base samples*

To better understand the effect of the reaction temperature on $H_{c2}$, we varied the A15 reaction temperature from 650 °C to 750 °C and determined the resulting $T_c$ values and $H_{c2}^{WHH}$ slope for the $x$Ta ($x$=2,4) base samples as summarized in Figure 10. In both cases, $T_c$ monotonically increases with increasing reaction temperature, whereas the $H_{c2}^{WHH}$ slopes either have a small variation or they decrease. These substantially opposite trends lead to a $H_{c2}^{WHH}(0)$ that is always better for the 4Ta case but with an almost constant value at 670 and 750 °C reaction temperatures for both 2 at% and 4 at% Ta doping levels.

*3. Effects of Ga addition*

Our experiments on the effect of including Ga in the Cu-Sn mixture did not reproduce earlier high claims and in fact revealed a significantly negative impact on both the A15 reaction temperature and $T_c$ values. By incorporating 20at% Ga into Cu-Sn powder mixture, we observed an onset $T_c$ of only 13 K, which also required the unreasonable reaction temperature of 1100 °C; $T_c$ values below 9.5 K were obtained for all heat treatments below 1100°C. To improve these results, we decreased the amount of Ga in the Cu-Sn powder mixture to 2at%, resulting in a lower A15 reaction temperature of 800-850 °C. The microstructures of the A15 reaction layer are shown in Figures 11a-b, clearly revealing the formation of Hf-rich precipitates at these high temperatures with an increase in their size and density as the A15 reaction temperature is increased. Our observations also revealed that the A15 grains in Ga-Ta-zHf wires coarsened immediately after the onset of A15 formation, while in Ta-Hf wires, the grain size was refined to as small as 70 nm [32]. These reaction temperatures are significantly higher than typically employed for magnets (650 °C – 670 °C), but they were necessary to obtain relatively homogeneous samples with good $T_c$. In Fig. 11c we compare $H_{c2}(T)$ for the 4Ta1Hf wires made with the with and without Ga and reacted at the same temperature of 850 °C. The two samples have comparable $H_{c2}^{WHH}(0)$ of 28.5-28.6 T. This is, however, 2 T below the maximum $H_{c2}^{WHH}(0)$ obtained on 4Ta1Hf without Ga at the much lower reaction temperature of 670°C.

## IV. Discussion

### A. Overview and value of a new $H_{c2}$ study

In the production of Nb$_3$Sn conductors, the addition of Ta and Ti has been a common practice since the 1980s [16,25,36]. The choice between using titanium or tantalum was originally the practical one of whether the commercial Nb0.8wt%Ti or Nb7.5wt.%Ta alloy was cheaper or more available. After it was found that replacing a few of the more widely available pure Nb filaments with standard Nb47wt.%Ti filaments enabled effective Ti doping throughout the wire [37,38], Ti doping largely replaced Ta as dopant in rod-based conductors like RRP, although Nb7.5wt.%Ta (Nb4Ta) was still favoured for powder in tube (PIT) conductors [9,39,40]. According to Suenaga both elements have a similar effect on raising $H_{c2}$ [16].

However, the new FCC target specification [41] explicitly links the need for the highest possible critical current density ($J_c$), which is encouraged by the smallest possible A15 grain size, to the need for the highest possible $H_{irr}$ and $H_{c2}$. This new perspective forces alloying strategy to consider both demands in addition to the internal oxidation approach [30,32,33].

This paper focuses on two aspects of the effect of alloying Nb with Ti, Ta, Zr and/or Hf, and the use of Ga: 1) the recrystallization behaviour of the Nb-alloy, which, if delayed above the normal A15 reaction temperature, sets up a denser nucleation of the growing A15 phase and 2) the upper critical field of the resulting Nb$_3$Sn A15 layers. We note at the outset that $H_{c2}$ alloying has been studied by many others, but generally only a few alloys at a time and often with limited access to high fields. This leads to difficulties in making reliable comparisons between studies and in understanding what properties are controlling changes in $H_{c2}$ as we discuss in the next section. The specific challenges of meeting the FCC specification are what prompted this systematic comparative study of alloys from one single source [32]. A common observation is that small alloy additions cause $H_{c2}$ to increase, likely because the alloying element creates disorder in the A15 compound [13,26]. The relationship between increased alloying and higher $H_{c2}$ may depend on whether the dopant enters the A15 lattice and which site it then occupies.

An essential study for explaining alloying effects in binary alloys is the thin film $H_{c2}$ study of Orlando *et al.* [10]. One view is that binary Nb$_3$Sn is always ordered [13,42], meaning that Sn-deficient A15 has the Nb chain sites always occupied by Nb atoms with vacant Sn sites as appropriate to the Sn content which can vary from 18-25at. %Sn. More recent calculations of the energetics of disorder in the A15 lattice by Besson *et al.* disagree with this view [43], concluding that anti-site disorder in sub-stoichiometric compositions is preferred. We do note that the Sn-deficient binary thin film made by Orlando *et al.* [10] with the highest $H_{c2}(0)$ of 29 T had a high resistivity of 36 $\mu\Omega cm$ and suppressed $T_c$ of 16.1 K, both suggestive of strong disorder.

In the alloyed A15 compositions of today's Ti and Ta-optimized wires, disorder is complicated because we must consider where Ta and Ti sit. To address the common finding that optimum, high-$J_c$ Ti-doped wires generally always had Sn-deficient A15 compositions (a maximum of ~23at. %Sn), EXAFS studies on commercial Ta, Ti, Ta + Ti RRP standard wires showed that Ta substitutes on both Nb and Sn sites but with a dependence on the heat treatment temperature, whereas Ti substitutes only on Nb sites independent of the reaction temperature [18,19]. However, this rather unexpected result is consistent with the result that Ti is about twice more effective (< 2 at. % Ti versus 4 at%. Ta) per at.% in raising $H_{c2}$. This implies that Ti preferentially produces disorder on the Nb A15 lattice sites, while Ta sitting on both Sn and Nb sites is a less potent disordering agent. Future EXAFS studies with bulk samples are planned to address this speculation in greater detail.

### B. Hardness, drawability, recrystallization of Nb-alloys and effect of alloying on A15 grain refinement

It was previously reported that Nb5Hf and Nb5Zr were sufficiently ductile for wire fabrication [27], a result which is confirmed here: in fact, we saw that all the Nb dopants work-hardened the alloy but not as much as is the case for the world's most heavily produced superconductor, Nb47wt.%Ti [44]. We have separately evaluated the ductility of the Nb4Ta1Hf alloy in commercially produced tubes, finding that it remains ductile for true drawing strains as high as 15 that enabled 22 mm diameter tubular filaments [44,45]. Work-hardening is generally most strongly affected by the size differential between the added solute and the matrix, Nb, which has a metallic atomic radius of 0.142 nm. The largest dopant was Hf (radius 0.160 nm) followed by Zr (0.160 nm), Ta (0.143 nm), and Ti (0.142 nm) respectively. In this respect the greater work-hardening exhibited by Zr and Hf is entirely expected.

Our observations indicate that the introduction of larger atoms like Hf and Zr and an increase in their concentration results in significant hardening of the Nb alloy, which is consistent with standard solid solution strengthening mechanisms. The hardness exhibits a near-linear increase with alloy content, as shown in Figure 2. Our Vickers hardness study revealed that the initial rate of increase in hardness is greater than that observed after a true-strain $\varepsilon = 3.5$. To evaluate the drawability and recrystallization temperature of the commercial Nb4Ta alloy and arc-melted Nb4Hf custom-alloy, both alloys were annealed at up to 1000°C for 3 hours after true strains of 7.0 and 8.1. The study shows that Nb4Hf increases the recrystallization temperature to ~800°C, in contrast to the Nb4Ta base alloy which was already recrystallized at 650°C, as is shown in Figure 3. The significance of adding Hf to increase the recrystallization temperature is emphasized by these findings.

Since unreacted Nb-alloy remained in the core following the A15 monofilament reaction, the FESEM-BSE image analysis showed that recrystallization of the Nb$x$Hf ($x$=1,3), Nb2Ta$x$Hf (Zr) ($x$=1,2), Nb2Ta($x$Hf$x$Zr) ($x$ = 0.5,1) alloys is always delayed until 750°C. On the other hand, Figure 5 shows that Nb$x$Ta ($x$=2,4), Nb4Ta0.8Ti, and Nb0.8Ti fully recrystallize at 750 °C. Incorporating the larger atoms of Hf or Zr, or a combination of both, into Ta does refine the A15 grain size during standard reactions [32,33]. It should be noted that the Nb-alloy recrystallization temperature increases with larger dopant atom sizes. Achieving this is extremely beneficial for A15 grain refinement, as the core alloy's grain structure directly affects the A15 nucleation. The higher recrystallization temperature enables more effective refinement of the A15 grain structure, resulting in better high-field performance of the Nb$_3$Sn conductor [32,33]. Qualitative analysis of FESEM-BSE images demonstrated that the addition of Hf (Zr) to Nb-$x$Ta ($x$=2,4) alloys refines the grain size of the A15 phase (Figure 5).

### C. WHH fit to estimate $H_{c2}$(T)

Figure 6 shows a good agreement between the experimental data and the WHH fits for data taken both up to 31 T and in our 16 T system. This agreement has a very important practical implication: Since measurements performed in fields up to 16 T (about 0.5 $H_{c2}$(0)) are sufficient to provide a good estimate of $H_{c2}$(0) without need for characterizations in high field facilities of limited access, a wide-range study of many alloys can be performed with limited field as here, where measurements in the NHMFL 31 T magnet was possible only for few samples. Moreover, similarly to what found by Orlando *et al.* for binary samples, we verified that $H_{c2}$ curves can be fitted without taking into account the Pauli paramagnetic limitation also for our ternary and quaternary A15 wires [46], as expected, considering that in all cases $H_{c2}$(0) does not exceed the zero-temperature BCS Pauli limit $\mu_0 H_{p,BCS}(0) = 1.86\, T_c$. For instance, the range of $H_{p,BCS}(0)$ for the alloyed samples listed in Table 1 (considering the $T_c^{WHH}$ values) is ~31.6-33.2 T, with $H_{p,BCS}$ up to 6.5 T larger than $H_{c2}$. The sample whose $H_{c2}$(0) most closely approaches $H_{p,BCS}(0)$ is the 4Ta1Hf wire heat treated at 670 °C, where the difference is reduced to only ~1.8 T. Moreover, in Nb$_3$Sn the effect of strong-coupling should probably be taken into account leading to an $H_p$ enhanced by a factor $\eta_{H_c}(1 + \lambda_{ep})^{1/2}$ with respect to the BCS value (from Orlando *et al.* [46]).

### D. Optimizing H$_{c2}$ by increasing disorder

The $H_{c2}$(0) of clean A15 wires is typically 25-26 T [47]: however, it is possible to reach values of up to 30 T in binary alloys using a variety of methods [48]. This suggests that disorder, likely in the Nb chain sites, is the main mechanism controlling $H_{c2}$. Earlier studies on bulk Nb-Sn samples have revealed that a normal state resistivity ($\rho_n$) value between 25 and 30 μΩ·cm can result in $H_{c2}$ values up to 30.0 T [10,48]. These values correspond to a $H_{c2}$ slope of 2.40 – 2.50 T/K and a $T_c$ range of 17.0 – 17.3 K. It is important to note that $T_c$ remains constant up to 20 μΩ·cm but drops rapidly above 30 μΩ·cm, which significantly degrades $H_{c2}$ [10]. Since the $H_{c2}$ slope is proportional to $\gamma \rho_n$, increasing disorder without losing too much $T_c$ is key to optimizing the $H_{c2}$ performance. Typically, the slope increases with $\rho_n$ up to 2.5 T/K, then decreases [10,46] and in Ta+Ti doped samples, the maximum $H_{c2}$ was obtained for a $\rho_n$ value of around 35 μΩ·cm [13]. Unfortunately, like almost all wires whether made in a laboratory as here or commercially, it is not possible to separately measure $\gamma$ and $\rho_n$. Therefore, here we focused on the influence of doping on the $H_{c2}$ slope and $T_c$, knowing that the optimal low-temperature performance is achievable by maximizing the first without losing too much $T_c$.

In our samples, we found that $T_c$ usually slightly increases with increasing reaction temperature, likely due to increased diffusivity of Sn into the growing A15 layer which becomes more homogeneous. However, this increased temperature is not beneficial for the disorder, as the $H_{c2}$ slope systematically decreases in all samples after increasing the reaction temperature to 750 °C. This is also true for the Nb4Ta alloy, which was investigated in a wide reaction temperature range. The combined opposite effects of the reaction temperature on $T_c$ and $H_{c2}$ slope leads to a mostly detrimental or negligible effect on $H_{c2}(0)$. We notice that this trend is different from that found in Ta-doped RRP [7,19] where increasing reaction temperature increases both $H_{c2}$ and the Kramer field (although with some weakening of the pinning performance). The different behaviours could be caused by the different wire designs, which could affect both the reaction path and the Sn diffusivity into the Nb alloy, affecting in turn the Ta occupancy. As verified by EXAFS studies, in RRP Ta occupies more Sn sites after a low temperature reaction, when the Sn supply is low, limiting the disorder on the Nb chains, whereas, when the Sn diffusivity is enhanced, Ta occupies mostly the Nb sites. Since in the present wires the Sn source is outside diffusing inward into the Nb alloy, the Sn supply in late stages of the reaction could be better than in RRP (where Sn diffuses outward), so leading to more Ta occupying the Nb sites already at a low reaction temperature.

Looking instead at the effect of the dopant concentration, we found that $T_c$ is suppressed with increasing Ta content in Ta-only alloyed wires, but with a corresponding increase of the low temperature $H_{c2}$: this is consistent with the behaviour previously observed by Suenaga [16], which also found the maximum $T_c$ at 2%Ta and the maximum $H_{c2}$ at ~4%Ta.

Additions of Ti or Hf by themself or together is instead not beneficial for the low-temperature $H_{c2}$ performance, indeed producing the worst $H_{c2}(0)$ values of our sample set. For Ti, this behaviour is different from earlier reports showing that 0.8Ti optimizes $H_{c2}$ [16]. The depressed $H_{c2}$ of our Ti-doped samples could be caused by the absence of Ti in the A15 structure. In fact, although in other Ti-doped Nb$_3$Sn samples Ti substitutes into the A15 structure, in our case the FESEM analysis reveals some dark-contrast areas (Figure 5c): Although they were too small to perform FESEM-EDS chemical characterization, the low BSE intensity is consistent with low atomic number Ti-rich precipitates. The worsening effect of Hf on $H_{c2}$ as the atomic% of Hf increases at 750 °C may be due to Hf precipitating out of the alloy (Figure 4) and not entering the A15 phase. Suenaga observed Hf-Sn precipitates in bronze-process wire and suggested this to be the primary reason for the relatively small benefit of Hf on $H_{c2}$ in comparison with Ta, Ti [16].

More interesting are the effects of combining different elements, like Ta, Hf and/or Zr. Although adding Hf or Zr to the Ta-base alloy is not beneficial for $T_c$ for the 2Ta, the double doping has a clear benefit in increasing the disorder as emphasized by the increase of the $H_{c2}$ slope. This leads to better $H_{c2}(0)$ in wires with a total doping percentage (Ta+Hf, or Ta+Zr), between 4 and 6% at 670 °C with the best performance observed in the 4Ta1Hf wire, followed closely by 4Ta1Zr, 4Ta2Hf and 2Ta2Hf. Figure 8c shows that the $H_{c2}$ slopes of $x$Ta and $x$Ta1Zr follow the same trend suggesting that the two elements similarly introduce disorder, possibly with similar occupancy, and only the total amount of dopants is relevant. On the other end the $H_{c2}$ slope trend for $x$Ta1Hf is clearly above that of $x$Ta and $x$Ta1Zr, suggesting that Hf and Ta synergistically increase disorder in the A15 phase. Interestingly, Ta has the same size as Nb, while Hf and Zr are larger, leaving it open for discussion whether, like their Group IVB sibling Ti, they also only occupy the Nb site.

Furthermore, the reaction temperature may also play a significant role in determining $H_{c2}$. In standard Ta- or Ti-doped RRP wires, higher $H_{c2}$ is always associated with higher reaction temperatures which make the A15 layer more Sn-rich [7]. However, in some of these new alloys, a lower temperature reaction yields higher $H_{c2}$, indicating that the relation between disorder and reaction temperature can also differ because of other factors. The lowering of $H_{c2}(0)$ by about 1 T for the highest $H_{c2}$ Hf and Zr alloys for reactions at 750 °C rather than 670 °C leads us to speculate that, like Ti, their group IVB siblings, Zr and Hf, also sit on the Nb sites but may be vulnerable to being pushed off this site by Ta switching from Sn to Nb sites with increasing reaction temperature. Despite the $H_{c2}(0)$ suppression observed for the reaction temperature of 750 °C with respect to 670 °C, also at high temperature the 4Ta1Hf wire produces a larger $H_{c2}$ slope than the other alloys.

We note that, despite the different efficiency, all these alloys introduce a significant amount of disorder as evidenced by the $H_{c2}$ slope reaching 2.54 T/K, compared to only 1.85 T/K for the binary system [47]. However, this increase in slope does not translate to a corresponding increase in $H_{c2}$, primarily due to the suppression of $T_c$.

We also re-examined the effect of Ga addition to the Cu-Sn powder mixture but together with the Nb4Ta1Hf alloy. However, introducing 2 at% Ga meant that a high A15 reaction temperature of 800 °C was required to form even a small amount of A15 (Figure 11). At a temperature of 850°C, the Ga-Hf samples underwent a full reaction, resulting in a complete annulus of A15 surrounding the core alloy but the high temperatures produced very large grains already at the onset of A15 formation as well as the formation of Hf precipitates. Comparable but depressed $H_{c2}$ values of 28.5-28.6 T are obtained in these conditions for the Ga-Ta-Hf and Ta-Hf wires, almost identical to those previously estimated for the Nb-Hf/Cu-Ga-Sn system [27,36]. It is worth noting that the addition of Ta to Hf resulted in an enhanced $H_{c2}(0)$, while the addition of Ta to Ga-Hf had minimal to no impact on $H_{c2}(0)$ values [27]. Previous studies by Bormann *et al.* [26] reported that $H_{c2}(0)$ could reach up to 33-35 T for 1.5-2.0 at% Ga addition to the binary A15 layer in co-evaporated thin film samples (no Ta or Hf). In our samples, even though FESEM-EDS analysis revealed the presence of 1.7 at% Ga in the A15 layer, the $H_{c2}$ value was low, consistent with its decreased $H_{c2}$ slope. In short no evidence was found in this study to support earlier reports of a positive influence of Ga addition on $H_{c2}$ [27].

$H_{c2}$ is only one of the properties that determined the high field performance of Nb$_3$Sn wires. Other superconducting characterizations (e.g. transport $J_c$) require the fabrication of multifilamentary, long and longitudinally uniform wires with a more suitable design and they will be studied in the future.

## V. CONCLUSIONS

In conclusion, we showed that the addition of up to 5 at% of Ta, Ti, Hf, Zr or a combination of these elements to Nb does not negatively impact their drawability. Using Hf or Zr by themselves or in combination with Ta or Ti can help to delay the recrystallization temperature and refine the grain size of the A15 phase. The optimization of $H_{c2}$ is temperature-sensitive and likely influenced by changes in site occupancy and disorder. The best approach to maximize $H_{c2}$ was to use a Nb4Ta base and add Hf or Zr, reaching a maximum $H_{c2}(0)$ of 30.2 T by using the Nb4Ta1Hf alloy and reacting the wire at 670 °C, closely followed by Nb4Ta1Zr leading to an $H_{c2}(0)$ of 30.0 T. Despite the $H_{c2}(0)$ enhancement, the WHH model without Pauli paramagnetic limiting can fit all our $H_{c2}(T)$ data, allowing us to characterize a variety of Nb alloys without need for high field measurements at 31 T. Unfortunately, the trend of the $H_{c2}$ slope and $T_c$ against A15 reaction temperature shows that it was not possible to simultaneously optimize these two parameters that determine $H_{c2}(0)$, producing almost constant $H_{c2}(0)$ in samples prepared with Nb$x$Ta ($x$=2,4 at%). Moreover, increasing the reaction temperature rapidly degrades the $H_{c2}$ performance of the Hf/Zr-doped samples. Our findings confirm the performance enhancing properties of Hf when combined to Nb4Ta, as recently reported by Balachandran *et al.*, without however providing further improvements. Contrary to earlier reports on Ga reaching $H_{c2}(0) \sim$ 33-35 T, we observed a strong depression of $H_{c2}(0)$ with as little as 2 at% Ga in Cu-Sn powder mixture. Despite Nb$_3$Sn being the first high field superconductor and remarkably still in wide use even 60 years after the first demonstration of its high field properties, it is clear that there is still much to learn about the limits of its high field properties.

## ACKNOWLEDGEMENTS

The authors extend their gratitude to Benjamin Walker (ASC-NHMFL) for early assistance with arc-melting, James Gillman (ASC-NHMFL) for his aid in wire fabrication, Van S Griffin (ASC-NHMFL) for his support during PPMS-16T measurements, and Eun Sang Choi (NHMFL) for his valuable contribution to the high field transport characterization. The research work was primarily funded by the grant number DE-SC0012083 from the US Department of Energy, Office of High Energy Physics. The study was conducted at the National High Magnetic Field Laboratory, which is sustained by the National Science Foundation's Cooperative Agreement No. DMR-2128556, as well as the State of Florida.

## AUTHOR ORCID ID

Nawaraj Paudel, 0000-0002-5482-5732
Chiara Tarantini, 0000-0002-3314-5906
Shreyas Balachandran, 0000-0002-0077-8504
Peter J. Lee, 0000-0002-8849-8995
David C. Larbalestier, 0000-0001-7098-7208